\documentstyle[epsfig,12pt]{article}
\begin{document}


\setcounter{page}{1}

\title{Search for Scalar Diquarks at the LHeC Based Gamma-Proton Collider}

\author{M.  \c{S}ahin$^a$ and O. \c{C}ak{\i}r$^b$ 
\\Physics Department, Faculty of Sciences, Ankara University,\\
06100, Tando\u{g}an, Ankara, Turkey\\
[5mm]
$^a$sahinm@science.ankara.edu.tr \\
$^b$ocakir@science.ankara.edu.tr\\}

\maketitle

\abstract{The diquark exotics which couple to a pair of quarks are
predicted by the Compositeness and Superstring inspired $E_{6}$
model. We study the single production of scalar diquarks at LHeC
based $\gamma p$ collider options. The background for three jet final
states are examined through appropriate kinematical cuts. We
discuss the possibility of measurements for the charges of scalar
diquarks ($qq$ and $qq'$).}

\section* {1. Introduction}

Diquarks are suggested by the models beyond the Standard Model (SM), such as
the superstring-inspired $E_{6}$ models \cite{Hewett89} and
composite models \cite{Terazawa80}. Diquarks have scalar and
vector form and carry baryon number $|B|=2/3$, and no lepton
number. They carry electric charges $|Q|=1/3$, $2/3$ or $4/3$.

Diquark production was examined in hadron colliders $p\bar{p}/pp$ \cite{CDF
Collaboration08,Atag99,Arik02,Cakir05,Mohapatra07,Sahin09,Han09}, $e^{-}e^{+}$
\cite{Cakir05,Gusso04} and $ep$ colliders \cite{Cakir05,Rizzo89}.
The collider dedector at Fermilab (CDF) has set limits on the masses of
scalar diquarks (predicted by $E_6$ model) decaying to dijets with the exclusion of mass
range $290<m_{DQ}<630$ GeV \cite{CDF Collaboration08}. This limit
is expected to be approximately valid for other scalar diquarks.
There is also indirect bounds imposed on couplings by the electroweak
precision data from LEP 
where these bounds allow diquark-quark cuplings up to a
value $\alpha_{DQ}$=$0.12$ \cite {Bhattacharyya95}.

In this work, we analysed single production of scalar diquarks in
a LHeC based $\gamma p$ collider. Interaction Lagrangian and
quantum numers of scalar diquarks are examined to calculate the decay widths, 
differential cross sections and total cross sections of the signal, and
the corresponding background. 
The signal and bacground analysis are performed 
for the scalar diquarks of $uu$, $ud$ and $dd$ types. 
The main parameters of the energy and luminosity options 
for a LHeC based $\gamma p$
collider are listed in table (1).

\begin{table}
\caption{Main parameters of energy and luminosity options of LHeC based $\gamma p$
collider.}
\label{table1}
\begin{tabular}{|c|c|c|c|c|c|}
\hline Collider & $E_{e}$(TeV) & $E_{p}$(TeV) &
$\sqrt{s_{ep}}$(TeV) & $\sqrt{s_{\gamma p}^{max}}$ & $L_{\gamma
p}^{int} L_{ep}^{int}(10^{2}pb^{-1})$\tabularnewline \hline \hline
LHeC & 0.07 & 7 & 1.40 & 1.28 & 10-100\tabularnewline \hline LHeC
& 0.14 & 7 & 1.98 & 1.80 & 10-100\tabularnewline \hline
\end{tabular}
\end{table}

\section* {2. Interaction Lagrangian}

Model independent, baryon number conserving, general
$SU(3)_{C} \times SU(2)_{W} \times U(1)_{Y}$ invariant effective
lagrangian for scalar and vector diquarks has the form
\cite{Atag99,Arik02}

\begin{eqnarray}
L_{|B|=2/3}=(g_{1L}\bar{q}^{c}_{L}i\tau_{2}q_{L}+g_{1R}\bar{u}^{c}_{R}d_{R})DQ^{c}_{1}
+\widetilde{g}_{1R}\bar{d}^{c}_{R}d_{R}\widetilde{DQ}^{c}_{1}
\nonumber \\
+\widetilde{g'}_{1R}\bar{u}^{c}_{R}u_{R}\widetilde{DQ'}^{c}_{1}+g_{3L}\bar{q}^{c}_{R}i\tau_{2}
\tau q_{L} \cdot \textbf{DQ}^{c}_{3}\hspace{0.6 in} \nonumber \\
+g_{2}\bar{q}^{c}_{L}\gamma_{\mu}d_{R}DQ^{c}_{2\mu}
+\widetilde{g}_{2}\bar{q}^{c}_{L}\gamma^{\mu}u_{R}\widetilde{DQ}^{c}_{2\mu}+H.c.\hspace{0.8
in}
\end{eqnarray}

In Eq. (1), $q_{L} = (u_{L}, d_{L})$ denotes the left-handed quark
spinor and $q^{c} = Cq^{T} (\bar{q}^{c} = -q^{T}C^{-1})$ is the
charge conjugated quark field. For the sake of simplicity, color
and generation indices are ommitted in (1). Scalar diquarks
$DQ_{1}$, $\widetilde{DQ}_{1}$, $\widetilde{DQ'}_{1}$ are
$SU(2)_{W}$ singlets and $\textbf{DQ}_{3}$ is a $SU(2)_{W}$
triplet. Vector diquarks $DQ_{2}$ and $\widetilde{DQ}_{2}$ are
$SU(2)_{W}$ doublets. At this stage, we assume that each SM
generation has its own diquarks and relevant couplings in order to
avoid flavour changing neutral currents. A general classification
of the first generation, color anti-triplet $(3^{*})$ diquarks is
shown in table 1 \cite{Cakir05}.

\begin{table}
\caption{Quantum numbers of the first generation, color anti-triplet
diquarks described by the effective lagrangian (1).}
\label{table2}
\begin{centering}
\begin{tabular}{cccccc}
 &  &  &  &  & \tabularnewline
\hline
 &  SU(3)$_{C}$ &  SU(2)$_{W}$ &  U(1)$_{Y}$ &  $Q$ &  Couplings \tabularnewline
\hline Scalar Diquarks  &  &  &  &  & \tabularnewline
 $DQ_{1}$ &  3$^{\star}$ &  1  &  2/3  &  1/3  &  $u_{L}d_{L}(g_{1L}),\, u_{R}d_{R}(g_{1R})$\tabularnewline
 $\widetilde{DQ}_{1}$ &  3$^{\star}$ &  1  &  -4/3  &  2/3  &  $d_{R}d_{R}(\tilde{g}_{1R})$\tabularnewline
 $\widetilde{DQ}_{1}^{\prime}$ &  3$^{\star}$ &  1  &  8/3  &  4/3  &  $u_{R}u_{R}(\tilde{g}_{1R}^{\prime})$\tabularnewline
 $DQ_{3}$ &  3$^{\star}$ &  3  &  2/3  &  $\left(\begin{array}{c}
4/3\\
1/3\\
-2/3\end{array}\right)$ &  $\left(\begin{array}{c}
u_{L}u_{L}(\sqrt{2}g_{3L})\\
\begin{array}{c}
u_{L}d_{L}(-g_{3L})\\
d_{L}d_{L}(-\sqrt{2}g_{3L})\end{array}\end{array}\right)$\tabularnewline
\hline Vector Diquarks  &  &  &  &  & \tabularnewline
 $DQ_{2\mu}$ &  3$^{\star}$ &  2  &  -1/3  &  $\left(\begin{array}{c}
1/3\\
-2/3\end{array}\right)$ &  $\left(\begin{array}{c}
d_{R}u_{L}(g_{2})\\
d_{R}d_{L}(-g_{2})\end{array}\right)$\tabularnewline
 $\widetilde{DQ}_{2\mu}$ &  3$^{\star}$ &  2  &  5/3  &  $\left(\begin{array}{c}
4/3\\
1/3\end{array}\right)$ &  $\left(\begin{array}{c}
u_{R}u_{L}(\tilde{g}_{2})\\
u_{R}d_{L}(-\tilde{g}_{2})\end{array}\right)$ \tabularnewline
\hline
\end{tabular}
\par\end{centering}
\end{table}

We consider the color $\textbf{3}^{*}$ scalar $DQ_{1}$ or
$DQ^{0}_{3}$ diquarks coupled to $ud$ pairs, $\widetilde{DQ}_{1}$
or $DQ^{-}_{3}$ diquarks coupled to $dd$ pair and
$\widetilde{DQ'}_{1}$ or $DQ^{+}_{3}$ diquarks coupled to $uu$
pair. Here, $DQ_3^{(+,0,-)}$ denotes the isospin triplet component of scalar diquarks. 
Diquark interactions with the gauge bosons are given as

\begin{equation}
L=\sum_{\Phi=DQ_{i}}(D_{\mu}\Phi)^{\dagger}(D^{\mu}\Phi)-M_{\Phi}^{2}\Phi^{\dagger}\Phi
\end{equation}

where covariant derivative is
$D_{\mu}=\partial_{\mu}-ig_{e}Q_{DQ}A_{\mu}-ig_{e}Q^{Z}Z_{\mu}-ig_{e}Q^{W}W_{\mu}-ig_{s}\frac{\lambda_{a}}{2}G_{\mu}^{a}$
where $A_{\mu}$, $Z_{\mu}$, $W_{\mu}$ and $G_{\mu}$denote photon,
$Z$- boson, $W-$boson and gluon fields, respectively. $Q_{DQ}$ is
the electromagnetic charge of a given diquark $DQ$ and
$Q^{Z}=(T_{3}-Q_{DQ}\sin^{2}\theta_{W})/(\cos\theta_{W}\sin\theta_{W})$
is the weak charge, $T_{3}$ the third component of the weak
isospin and $\theta_{W}$ is the Weinberg angle, $g_{s}$ is the
strong coupling constant and $\lambda_{a}$ are the Gell-Mann
matrices.

The decay widths ($\Gamma_{DQ}$) for scalar diquark is calculated from 
the equation (1) and we plot the diquark decay width 
versus diquark mass in fig. 1.

\begin{figure}[!htb]
\epsfig{figure=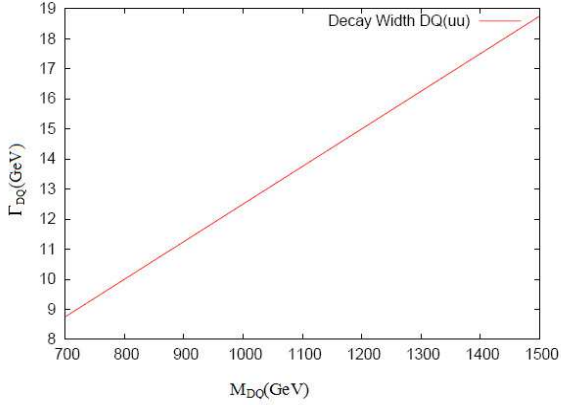 ,width=8cm,height=6cm,angle=0}
\caption{Decay width versus mass for scalar diquark $DQ(uu)$.\label{fig:figure1}}
\end{figure}

\section* {3. Production Cross Section for Scalar Diquarks}

Scalar diquarks can be produced singly via the subprocess $\gamma$
$p$$\rightarrow$$q^{'}DQ$ and the differential cross section is given
by

\begin{eqnarray}
\frac{d\hat{\sigma_{S}}}{d\widehat{t}}=\frac{g_{e}^{2}g_{DQ}^{2}}{16\pi\widehat{s}^{2}}\left[\frac{Q^{\prime2}\widehat{s}}{(\widehat{s}+\widehat{t}-m_{DQ}^{2})}-\frac{2QQ^{\prime}(\widehat{s}+\widehat{t})(\widehat{s}-m_{DQ}^{2})}{\widehat{s}(\widehat{s}+\widehat{t}-m_{DQ}^{2})}\right.
\nonumber \\ +\frac{Q^{2}(\widehat{s}+\widehat{t}-m_{DQ}^{2})}{\widehat{s}}-\frac{Q_{DQ}Q\widehat{t}(\widehat{s}-2m_{DQ}^{2})(\widehat{t}-m_{DQ}^{2})}{\widehat{s}(\widehat{t}^{2}+\Gamma_{DQ}^{2}m_{DQ}^{2}-2\widehat{t}m_{DQ}^{2}+m_{DQ}^{4})}\nonumber \\
 \left.+\frac{Q_{DQ}^{2}\widehat{t}(\widehat{t}+m_{DQ}^{2})}{(\widehat{t}^{2}+\Gamma_{DQ}^{2}m_{DQ}^{2}-2\widehat{t}m_{DQ}^{2}+m_{DQ}^{4})} \right.
 \nonumber  \\
\left. + \frac{Q_{DQ}Q^{\prime}\widehat{t}(\widehat{t}-m_{DQ}^{2})((\widehat{s}+\widehat{t}+m_{DQ}^{2}))}
{(\widehat{s}+\widehat{t}-m_{DQ}^{2})(\widehat{t}^{2}+\Gamma_{DQ}^{2}m_{DQ}^{2}-2\widehat{t}m_{DQ}^{2}+m_{DQ}^{4})}
 \right]
\end{eqnarray}

where $\hat{s}$ and $\widehat{t}$ are Mandelstam variables for
the subprocess $\gamma q\to DQ q'$. $Q$ and $Q^{\prime}$ are the charges of initial
and final quarks, respectively. As it can be seen from equation (3), the cross 
section is proportional to diquark charges. Therefore, the 
diquark charges can be identified at a LHeC based $\gamma$p collider.
In figure 2, the hadronic process for diquark single production is
shown.

\begin{figure}[!tb]
\centering \epsfig{figure=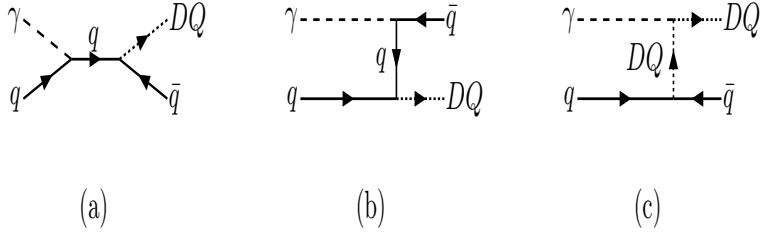
,width=10cm,height=3cm,angle=0} \caption{Diagrams for the single
production of scalar diquarks at LHeC based $\gamma p$ collider.\label{fig:figure2}}
\end{figure}

The signal for diquark single production would clearly manifest
itself in three jets cross sections. The total cross section for
the single production of diquarks at $\gamma$p collider is given
by

\begin{equation}
\sigma_{DQ}^{S}=\sum_{q}\int_{M_{DQ}^{2}/s}^{0.83}d\tau\int_{\tau/0.83}^{1}\frac{dx}{x}f_{q/p}(x,Q^{2})f_{\gamma/e}(\frac{\tau}{x})\int_{\widehat{t}_{\min}}^{\widehat{t}_{\max}}d\widehat{t}\frac{d\hat{\sigma}_{DQ}^{S}}{d\hat{t}}
\end{equation}

where $f_{q}(x,Q^{2})$ is the quark distribution functions from
the proton. The third integration over $\hat{t}$ is taken in the
interval $\hat{t}_{\min}$ and $\hat{t}_{\max}$, where
$\hat{t}_{\min}=-\hat{s}+m_{DQ}^{2}$ and $\hat{t}_{\max}=0$. The
energy spectrum of the Compton backscattered photons from
electrons is given by \begin{equation}
f_{\gamma/e}(z)=\frac{1}{D(\xi)}\left[1-z+\frac{1}{1-z}-\frac{4z}{\xi(1-z)}+\frac{4z^{2}}{\xi^{2}(1-z)^{2}}\right],\end{equation}
with

\begin{equation}
D(\xi)=(\frac{1}{2}+\frac{8}{\xi}-\frac{1}{2(1+\xi)^{2}})+(1-\frac{4}{\xi}-\frac{8}{\xi^{2}})\ln(1+\xi),\end{equation}

where $\xi=4E_{e}\omega_{0}/m_{e}^{2},$ and $z=E_{\gamma}/E_{e}$ is
the ratio of the backscattered photon energy to the initial
electron energy. The energy $E_{\gamma}$ of converted photons
restricted by the condition $z_{\max}=0.83$. The value
$z_{\max}=\xi/(\xi+1)=0.83$ corresponds to $\xi=4.8$ as given in
\cite{Ginzburg89}.

In figure 3, total cross sections for scalar diquarks depending on
the electron beam energies are shown. From these plots we see the
high energy and low energy behaviour of the total cross section
for a given value of $m_{DQ}$ and $\alpha_{DQ}$. The total cross
sections have no divergencies at large energies. Thus, equation (4) prove
the unitarity condition. In figure (4), the total cross sections versus
scalar diquark masses are plotted for the LHeC
($\sqrt{s}=1.4$ TeV) energy with the coupling $\alpha_{DQ}=0.1$
using CTEQ parton distribution functions \cite{cteq5l00} at the
factorization scale $Q^{2}=m_{DQ}^{2}$. From these figures we find
that diquarks with charge $|Q| = 4/3$ have the largest cross
sections when compared to the other types.

\begin{figure}[!tb]
\centering \epsfig{figure=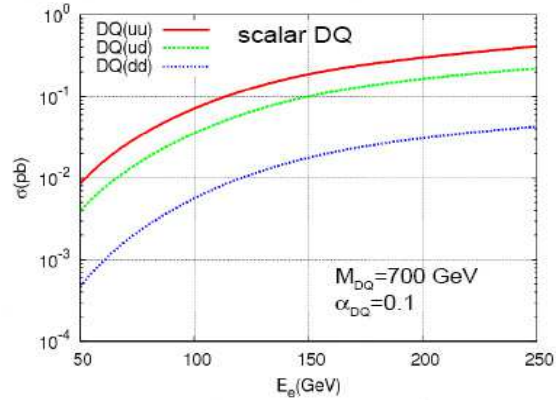
,width=8cm,height=6cm,angle=0} \caption{Total cross sections for
single scalar diquark ($m_{DQ}=700$ GeV) production as a function
of the electron beam energy ($E_{e}=50-250$ GeV) at $ep$
colliders, where proton beam (from the LHC) has an energy
$E_{p}=7000$ GeV.} \label{figure3}
\end{figure}

\begin{figure}[!tb]
\centering \epsfig{figure=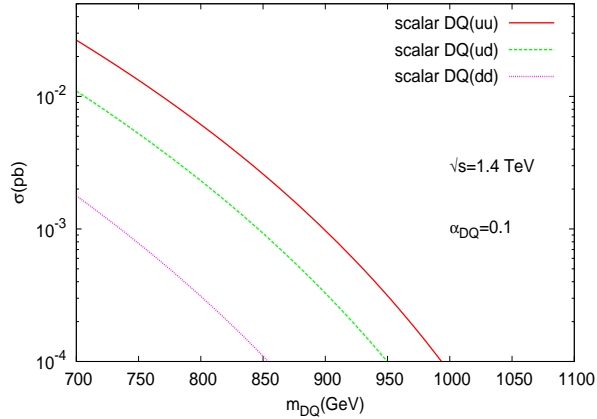
,width=8cm,height=6cm,angle=0} \caption{Total cross sections as a
function of scalar diquark masses at LHeC with $\sqrt{s}=1.4$
TeV.} \label{figure4}
\end{figure}

\section* {4. Signal and Background}

We generate diquark signal and the corresponding background events with
the program CalcHEP \cite{Pukhov99}. Here, we consider two types of background 
one is interfering with the signal events and the other is 
reducible background contributing to three-jet events. 
The background for three-jet events have large cross sections, 
since the signal has different shape than the
background, still we have opportunity to reduce these 
backgrounds by applying suitable
kinematical cuts.  In figure (5), the $p_{T}$ distribution of 
3 jets for signal with $M_{DQ}=700$ GeV are shown. The jets 
from diquarks have large transverse momentum distribution 
around the half value of the diquark mass. 
Thus, we need at least 20 GeV for the transverse momentum cut and 
additional kinematics variables to reduce background more efficiently. 
In figure (6), the pseudo-rapidity distribution for signal with
$M_{DQ}=700$ GeV and background are shown at LHeC with
$\sqrt{s}=1.4$ TeV.

\begin{figure}[!tb]
\centering \epsfig{figure=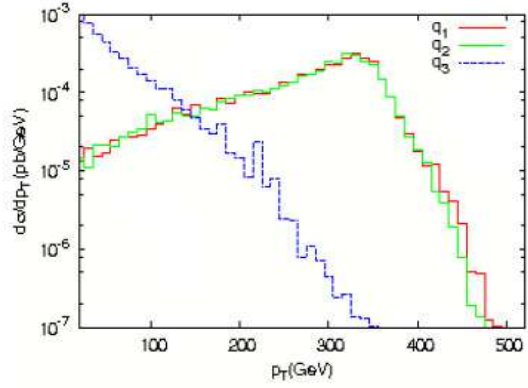
,width=8cm,height=6cm,angle=0} \caption{Transverse momentum
distributon for scalar diquarks with $M_{DQ}=700$ GeV at
$\sqrt{s}=1.4$ TeV LHeC energy options.} \label{figure5}
\end{figure}

\begin{figure}[!tb]
\centering \epsfig{figure=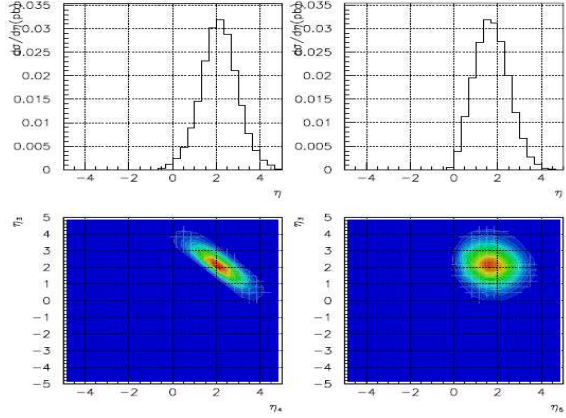
,width=8cm,height=6cm,angle=0} \caption{Pseudo-rapidity
distribution for the jets from scalar diquark production with $M_{DQ}=700$ GeV at
$\sqrt{s}=1.4$ TeV LHeC energy option.} \label{figure6}
\end{figure}

From figure (6), signal jets from scalar diquark are mostly located in the 
pseudo-rapidity ($\eta$) region 
$1 < \eta < 3.2$. In figure (7), one can see 
the distribution of pseudo-rapidity (mostly in the range $1 < \eta <2$) from SM three-jet background 
at LHeC with $\sqrt{s}=1.4$ TeV. The invariant mass distribution of dijets from the scalar
diquark signal and the SM background are shown in figure (8).

\begin{figure}[!tb]
\centering \epsfig{figure=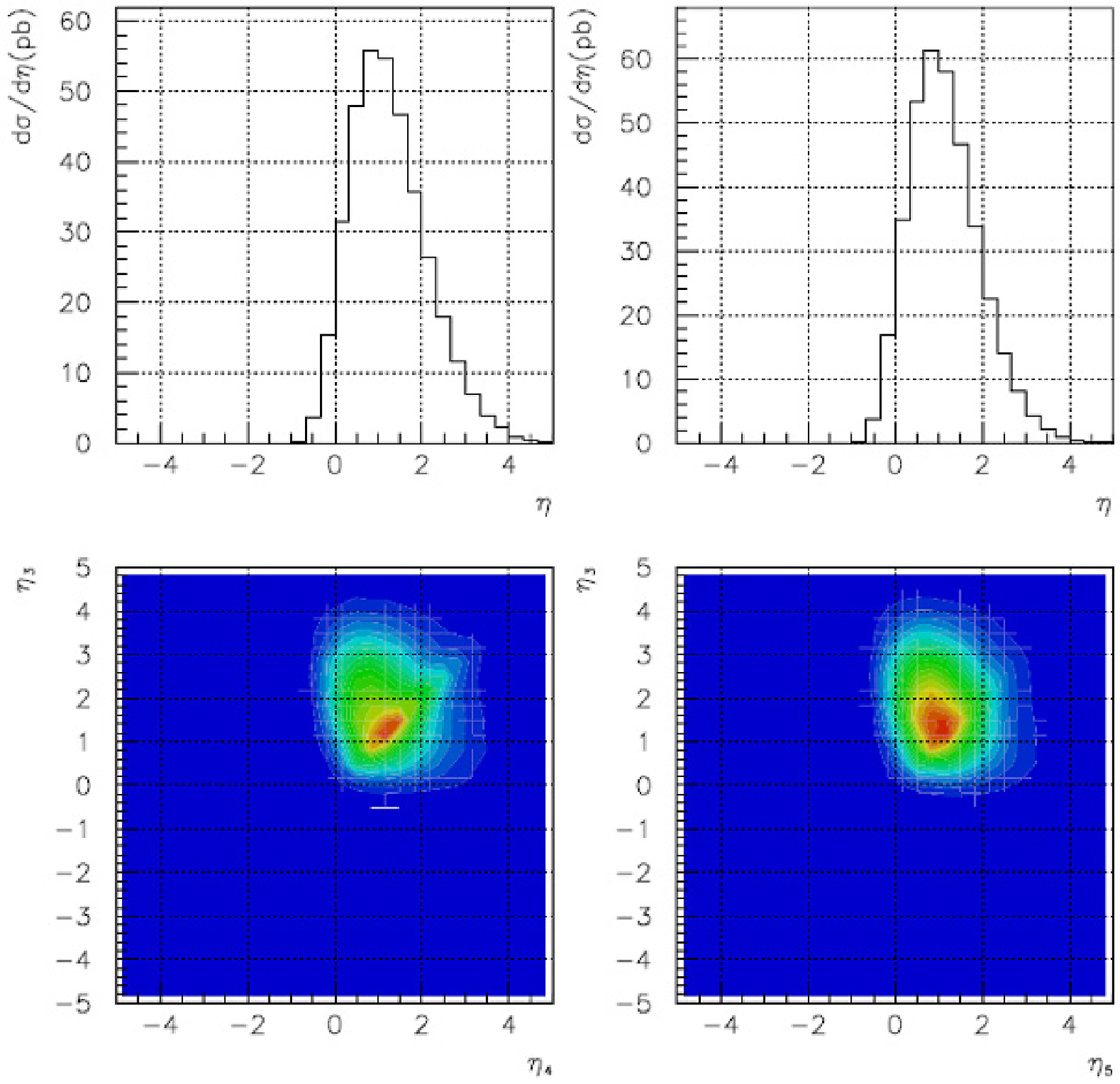
,width=8cm,height=6cm,angle=0} \caption{Pseudo-rapidity
distributions for SM $3jet$ background at $\sqrt{s}=1.4$ TeV LHeC
energy option.} \label{figure6}
\end{figure}

\begin{figure}[!tb]
\centering \epsfig{figure=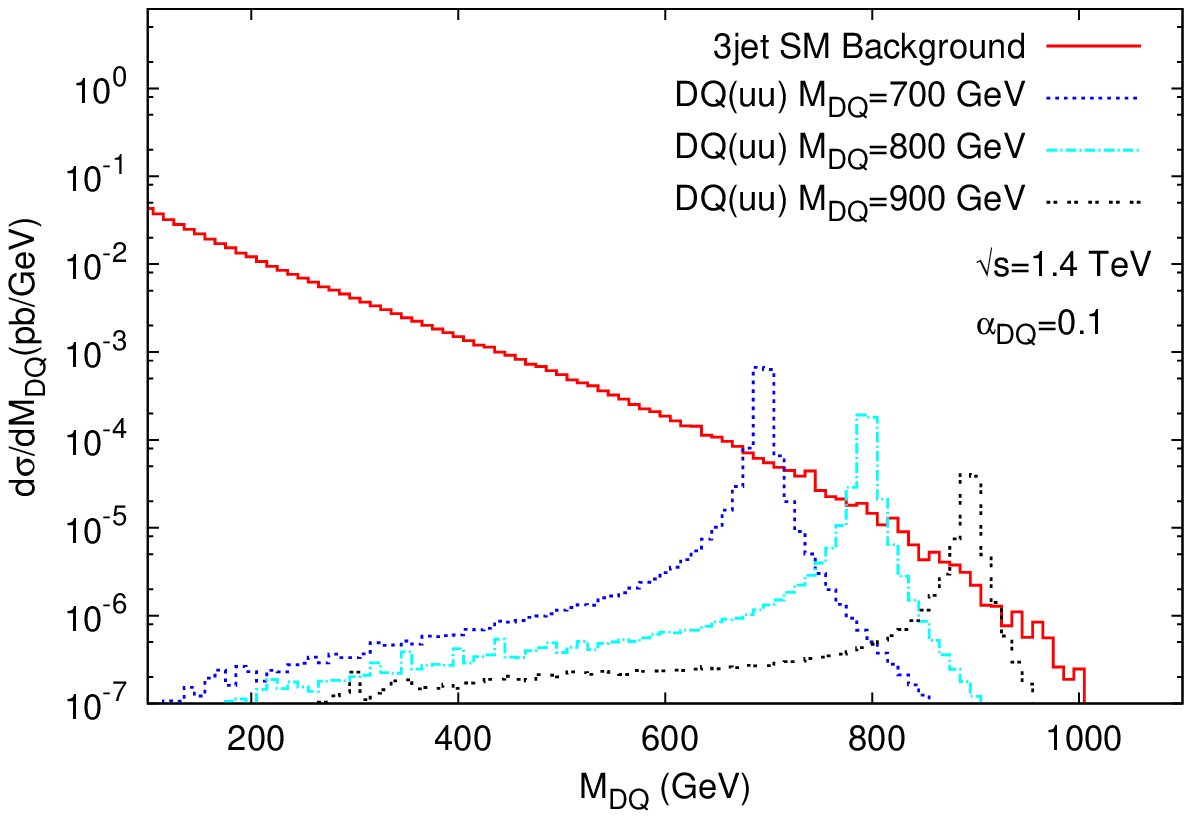
,width=8cm,height=6cm,angle=0} \caption{Dijet invariant mass
distributions for $ep\rightarrow DQjX\rightarrow jjj$. Resonance
peaks are shown for scalar and vector diquark masses 700, 800, 900
GeV for comparison with smooth QCD backgrounds.} \label{figure8}
\end{figure}

From these figures, more appropriate cuts for signal jets are $1
<\eta^{j} < 3.2$, $m_{DQ}-\Delta{m} < m_{jj} < m_{DQ} +
\Delta{m}$, $p^{j}_{T}>20$ GeV at LHeC $\sqrt{s}=1.4$ TeV energy option. Same
calculations have been performed for LHeC $\sqrt{s}=1.96$ TeV energy option.
In this case, the appropriate cuts for signal jets has been founded as $-0.5
<\eta^{j} < 3$, $m_{DQ}-\Delta{m} < m_{jj} < m_{DQ} + \Delta{m}$,
$p^{j}_{T}>20$ GeV at
LHeC $\sqrt{s}=1.96$ TeV energy option.

In order to obtain the observability of diquarks at LHeC based gamma-p 
collider we have
calculated the signal (S) and background (B) event estimations for an
integrated luminosity of $10^{4}$ $pb^{-1}$ for
one year of operation. The signal generated by a diquark of mass
$m_{DQ}$ and decay rate $\Gamma_{DQ}$ is calculated integrating
the differential cross section in the two-jet invariant mass
interval $m_{DQ} - \Gamma_{DQ} < m_{jj} < m_{DQ} + \Gamma_{DQ}$
which gives approximately $95\%$ of the events around the
resonance. For a realistic analysis of the background events we
take into account the finite energy resolution of the generic
hadronic calorimeter as $\delta{E}/E = 0.5/\sqrt{E}
+ 0.03$ for jets with $|y| < 3$. The corresponding two-jet
invariant mass resolution is given approximately by
$\delta{m_{jj}} = 0.5\sqrt{m_{jj}} + 0.02m_{jj}$. The background
is calculated by integrating the cross sections in the range
$m_{DQ}-\Delta{m} < m_{jj} < m_{DQ} + \Delta{m}$ with $\Delta{m} =
max(\Gamma_{DQ}, \delta{m_{jj}})$. The significance of signal over
the background is defined as $S/\sqrt{B}$. Thus we used appropriate
cuts and the dedector parameters to find the observability of
diquarks at LHeC based $\gamma p$ collider. 
Then, we listed the values in table (3) and (4) where
SS represent significance of diquarks.

\begin{table}
\caption{Observability of diquarks at $\gamma p$ collider based on LHeC $\sqrt{s}=1.4$ TeV.}
\label{table3}
\begin{tabular}{|c|c|c|c|}
\hline 
 & \multicolumn{3}{c|}{$SS$ for $L_{int}=10^{4}pb^{-1}$}\tabularnewline
\hline 
$M_{DQ}$(GeV)  & $DQ(uu)$  & $DQ(ud)$  & $DQ(dd)$\tabularnewline
\hline 
700  & 17.2  & 3.6  & 1.2\tabularnewline
\hline 
800  & 10.1  & 1.8  & ---\tabularnewline
\hline 
900  & 4.8  & ---  & ---\tabularnewline
\hline
\end{tabular}
\end{table}

\begin{table}
\caption{Observability of diquarks at $\gamma p$ collider based on LHeC $\sqrt{s}=1.98$ TeV.}
\label{table4}
\begin{tabular}{|c|c|c|c|}
\hline 
 & \multicolumn{3}{c|}{$SS$ for $L_{int}=10^{4}pb^{-1}$}\tabularnewline
\hline 
$M_{DQ}$(GeV)  & $DQ(uu)$  & $DQ(ud)$  & $DQ(dd)$\tabularnewline
\hline 
700  & 36.4  & 10.8  & 3.6\tabularnewline
\hline 
800  & 28.3  & 7.8  & 2.5\tabularnewline
\hline 
900  & 19.9  & 5.1  & 1.6\tabularnewline
\hline 
1000  & 14.9  & 3.4  & 1.0\tabularnewline
\hline 
1100  & 10.4  & 2.1  & ---\tabularnewline
\hline 
1200  & 6.3  & ---  & ---\tabularnewline
\hline
\end{tabular}
\end{table}

If we take at least $10$ signal events and $S/\sqrt{B}\geq 3$ as
observability criteria, For the diquarks with charge $|Q| = 2/3$ it is
possible to cover mass ranges up to $0.8$ TeV at the LHeC with
$L_{int} = 10^{4} pb^{-1}$ and $\sqrt{s}=1.98$ TeV.
The scalar diquarks with charge $|Q| = 4/3$ can be observed
up to $1.2$ TeV at $\sqrt{s}=1.98$ TeV.

\section*{5. Conclusion}

If diquarks exist, LHC could find them in resonance channel,
however their charges and coupling types can be identified at a
LHeC based $\gamma p$ collider. Up to 1.2 TeV mass of diquarks can
be studied at LHeC based $\gamma p$ collider. In the
single production mechanism, the spin of the diquarks 
can also be determined by studying the angular distributions of 
the final state jets with high $p_T$. 

\section*{Acknowledgements}

This work is supported in part by the Turkish Atomic Energy Authority (TAEK) and the State Planning Organization (DPT) with grant number DPT2006K-120470.


\begin{thebibliography}{25}

\bibitem{Hewett89} J. L. Hewett and T. G. Rizzo, Phys. Rep., {\bf 183}, 193, 1989.\\[-7mm]
\bibitem{Terazawa80}H. Terazawa, Phys. Rev. {\bf D22}, 184, 1980.\\[-7mm]
\bibitem{CDF Collaboration08} CDF Collaboration, CDF note {\bf 9246},2008. \\[-7mm]
\bibitem{Atag99}S. Atag, O. Cakir, and S. Sultansoy,Phys. Rev., {\bf D59}, 015008, 1999.\\[-7mm]
\bibitem{Arik02} E. Arik, S. A. Cetin, O. Cakir and S. Sultansoy, J. High
Energy Phys., {\bf 09}, 024, 2002.\\[-7mm]
\bibitem{Cakir05}O. Cakir, and M. Sahin, Phys. Rev., {\bf D72}, 115011, 2005.\\[-7mm]
\bibitem{Mohapatra07}R.N.Mohapatra, Nobuchika Okada, and Hai-Bo
Yu, Phys.Rev. D {\bf 77}, 01170(R), 2008.\\[-7mm] 
\bibitem{Sahin09}M. Sahin, and O. Cakir, Balkan Physics Letters (BPL), {\bf16(1)}, pp. 120-125, 161020 (2009).\\[-7mm]
\bibitem{Han09}Tao Han, Ian Lewis, and Thomas McElmurry, arXiv:0909.2666v1 [hep-ph] 15 Sep 2009.\\[-7mm]
\bibitem{Gusso04}A. Gusso, J. Phys. G:Nucl. Part. Phys. 30 (2004) 691-702.\\[-7mm]
\bibitem{Rizzo89} T.G. Rizzo, Z.Phys. C {\bf 43}, 223, 1989.\\[-7mm]
\bibitem{Bhattacharyya95} G. Bhattacharyya, D. Choudury and K. Sridhar,Phys. Lett., {\bf B355},193, 1995.\\[-7mm]
\bibitem{Ginzburg89} G. Bhattacharyya, D. Choudury and K. Sridhar,Phys. Lett., {\bf B355},193, 1995.\\[-7mm]
\bibitem{cteq5l00}CTEQ Collaboration, H.L. Lai et al., Eur.Phys. J.\textbf{C12} (2000) 375.
\bibitem{Pukhov99} A.Pukhov et al.,{\bf hep-ph/9908288}; A. Pukhov, e-Print Archive, {\bf hep-ph/0412191}, 2004.\\[-7mm]
\bibitem{Schrempp86} B. Schrempp, {\bf MPI-PAE/PTh}, 72-86, 1986.; W. Buchmuller, Acta Phys. Austriaca,{\bf 27}, 517,1985\\[-7mm]
\bibitem{Lai00} H. L. Lai et al. (CTEQ Collaboration), Eur. Phys. J., {\bf C 12}, 375, 2000.\\[-7mm]
\bibitem{Atlas99} ATLAS Collaboration, Report No. ATLAS TDR 14,
CERN/LHCC 99-14, 1999; Report No. ATLAS TDR 15, CERN/LHCC 99-15,
1999.\\[-7mm]

\end{thebibliography}
\end{document}